%% file: main.tex
\documentclass[a4paper,conference]{IEEEtran}

 \input{Setting2}

\begin{document}

\title{DNN-Alias: Deep Neural Network Protection Against Side-Channel Attacks via Layer Balancing }
\author{
 \IEEEauthorblockN{Mahya~Morid~Ahmadi$^\dag$,
        Lilas~Alrahis$^\ddag$,
       Ozgur~Sinanoglu$^\ddag$ and Muhammad~Shafique$^\ddag$}
  	\IEEEauthorblockA{$^\dag$\textit{Technische Universit\"at Wien (TU Wien), Vienna, Austria}\\
  	$^\ddag$\textit{Division of Engineering, New York University Abu Dhabi (NYUAD), Abu Dhabi, United Arab Emirates}\\
  	Email: \ mahya.ahmadi@tuwien.ac.at, \{lma387, ozgursin, muhammad.shafique\}@nyu.edu}\\ \vspace{-20pt}
 }

\maketitle

\begin{abstract} 
Extracting the architecture of layers of a given deep neural network (DNN) through hardware-based side channels allows adversaries to steal its intellectual property and even launch powerful adversarial attacks on the target system. In this work, we propose \textit{DNN-Alias}, an obfuscation method for DNNs that forces all the layers in a given network to have similar execution traces, preventing attack models from differentiating between the layers. Towards this, DNN-Alias performs various layer-obfuscation operations, e.g., layer branching, layer deepening, etc, to alter the run-time traces while maintaining the functionality. DNN-Alias deploys an evolutionary algorithm to find the best combination of obfuscation operations in terms of maximizing the security level while maintaining a user-provided latency overhead budget. 

We demonstrate the effectiveness of our DNN-Alias technique by obfuscating the architecture of 700 randomly generated and obfuscated DNNs running on multiple Nvidia RTX 2080 TI GPU-based machines. Our experiments show that state-of-the-art side-channel architecture stealing attacks cannot extract the original DNN accurately. 
 Moreover, we obfuscate the architecture of various DNNs, such as the VGG-11, VGG-13, ResNet-20, and ResNet-32 networks. Training the DNNs using the standard CIFAR10 dataset, we show that our DNN-Alias maintains the functionality of the original DNNs by preserving the original inference accuracy. Further, the experiments highlight that adversarial attack on obfuscated DNNs is unsuccessful.

\end{abstract}
\vspace{-0.51em}


\input{Text/01_Introduction.tex}
\input{Text/02_Background}
\input{Text/04_Methodology}

\input{Text/05_ExperimentalSetup}

\input{Text/06_Results}

\input{Text/07_Conclusion}

\bibliographystyle{IEEEtran}
\bibliography{main}

\end{document}

%% file: Setting2.tex
\IEEEoverridecommandlockouts
\usepackage{enumitem}
\usepackage{comment}

\usepackage{cite}

\usepackage{multirow}
\usepackage{amsmath,amssymb,amsfonts}
\usepackage{graphicx}
\usepackage{textcomp}
\usepackage{stmaryrd}
\usepackage{xcolor}
\usepackage{tabularx,booktabs}

\def\BibTeX{{\rm B\kern-.05em{\sc i\kern-.025em b}\kern-.08em
    T\kern-.1667em\lower.7ex\hbox{E}\kern-.125emX}}
\newcommand{\red}[1]{\textcolor{red}{#1}}
\newcommand{\blue}[1]{\textcolor{blue}{#1}}

\input{math_commands}

\usepackage{circledsteps}
\pgfkeys{/csteps/inner color=white}
\pgfkeys{/csteps/outer color=none}
\pgfkeys{/csteps/fill color=black}
\usepackage{pifont} 
\newcommand{\cmark}{\ding{51}}%
\newcommand{\xmark}{\ding{55}}
\definecolor{cadmiumgreen}{rgb}{0.0, 0.42, 0.24}
\usepackage{setspace}
\usepackage{algorithm}
\usepackage[%
    indLines=false,
    italicComments=true,
    noEnd=false,
]{algpseudocodex}

%% file: math_commands.tex
\usepackage{amsmath,amsfonts,bm}

\def\eqref#1{equation~\ref{#1}}

\def\1{\bm{1}}

\def\mA{{\bm{A}}}
\def\mB{{\bm{B}}}

\def\mI{{\bm{I}}}

\def\mU{{\bm{U}}}
\def\mV{{\bm{V}}}
\def\mW{{\bm{W}}}
\def\mX{{\bm{X}}}

\DeclareMathAlphabet{\mathsfit}{\encodingdefault}{\sfdefault}{m}{sl}
\SetMathAlphabet{\mathsfit}{bold}{\encodingdefault}{\sfdefault}{bx}{n}



%% file: Text/01_Introduction.tex
\section{Introduction}
\vspace{-0.47em}
Deep neural networks (DNNs) have experienced rapid advancements in the past decade, leading to their application to many areas of human endeavor and various fields of science~\cite{tesla,siri}. The deployment of DNNs in mission-critical applications, such as healthcare systems~\cite{healthcare} and anomaly detection in cyber-physical systems~\cite{anomaly}, raises concerns about the safety and security of these networks. 
Moreover, building and training DNNs require expert knowledge and costly resources. Thus, DNNs for a certain application are considered sensitive and expensive intellectual property (IP) that require protection from malicious users and/or market competitors.

To facilitate the application of DNNs, model developers offer machine learning as a service (MLaaS), which includes machine learning (ML) models and services running on the cloud or edge devices~\cite{mlaas_Sec}. In the MLaaS business model, the end-user receives input/output access to the DNN, i.e., \textit{black-box access}. An untrusted user would want to extract the IP of the underlying DNN, for launching \textit{white-box adversarial attacks} on the target system and/or for stealing the IP without incurring high research and development costs.\footnote{An adversarial attack performs subtle perturbations to the input samples of an ML model, causing the model to predict incorrect outputs~\cite{AdversarialAttack,fadec}. A white-box attack model assumes access to the inputs, architecture, and internal details, e.g., weights. Conversely, in the black-box model, the attacker lacks access to these details. A gray-box attack trains a substitute model to generate adversarial samples and attack the target model~\cite{grey}.} Note, it is difficult to launch such attacks with only back-box access to the system, e.g., it takes 30 GPU days to launch an adversarial attack on Lenet+ (7 layers) using input/output queries~\cite{iclr2018}.

\begin{figure}[!t]
 \centering
 \includegraphics[width=\linewidth]{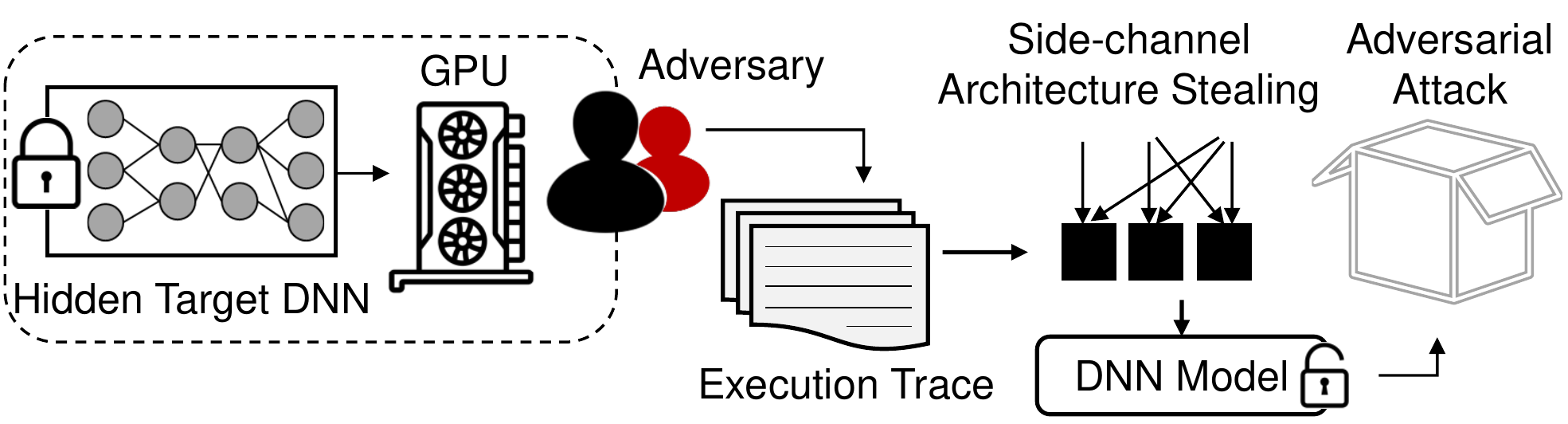}
  \vspace{-1.2em}
 \caption{Hardware-based side-channel leakage facilities DNN architecture stealing, leading to white-/grey-box adversarial attacks.}
  \vspace{-1.2em}
 \label{fig:ModelStealing}
\end{figure}

Additional information about the target DNN could be extracted through \textit{hardware-based side-channels}, such as the DNN architecture (i.e., the number, type, dimension, and connectivity of layers), enabling advanced adversarial attacks. For example, \textit{DeepSniffer}~\cite{DeepSniffer} is a DNN side-channel-based architecture stealing (SCAS) attack that learns the correlation between the architecture hints (such as volumes of memory writes/reads) and the 
DNN architecture.\footnote{SCAS attacks can be physical (edge) or remote (cloud)~\cite{remote}.} DeepSniffer showed that including the architecture information increases the success rate of an adversarial attack by $\approx3\times$. Fig.~\ref{fig:ModelStealing} shows an example of such SCAS attack flow.

\begin{figure}[!t]
 \centering
 \includegraphics[width=\linewidth]{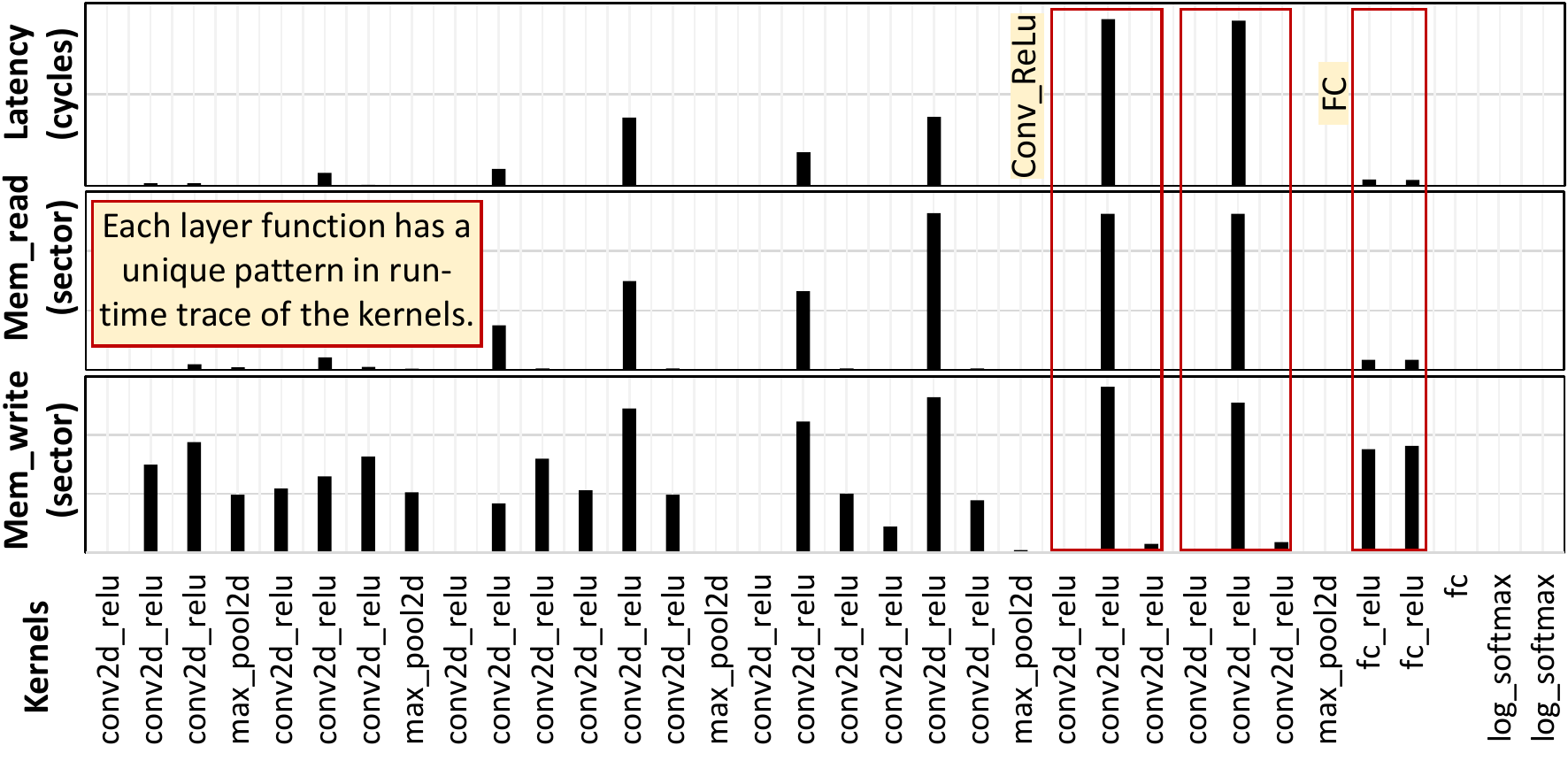}
 \vspace{-1.2em}
\caption{
The run-time trace of AlexNet. The unique pattern of the memory access bytes and computational cycles per each kernel allows SCAS attacks to learn these features and extract the target DNN architecture successfully.}
 \vspace{-1.2em}
\label{fig:Case_Study}
\end{figure}

To demonstrate the potent nature of such SCAS attacks, we plot the run-time traces of the AlexNet DNN trained on the CIFAR10 dataset in terms of computation latency (cycles) and memory access time (read and write) per layer in Fig~\ref{fig:Case_Study}. It can be observed that each type of layer has a unique execution signature. As a result, SCAS attacks learn the execution pattern of the layer functions and extract the layer sequence.

Researchers have studied SCAS attacks based on various side-channels, e.g., power~\cite{Power}. We focus on timing (execution time and memory access) side-channels, which are more stealthy, as they exploit an operational part of system (system profilers) for measuring and do not require extra equipment.

Researchers have developed several protection techniques to thwart SCAS attacks. Some \textit{hiding} techniques 
decrease the signal-to-noise ratio in side-channel traces, e.g., via dummy memory operations~\cite{Parasite}. Other methods necessitate hardware modifications to encrypt memory and other side-channel leakage~\cite{shi2011oblivious}. More recently, DNN obfuscation has been proposed to alter the run-time traces of a given DNN while preserving its functionality, thwarting SCAS attacks~\cite{li2021neurobfuscator,dnncloak}. Nevertheless, these methods suffer at least from one of the drawbacks discussed next and summarized in Table~\ref{tab:comparison}.

\begin{table}[!t]
\centering
\vspace{-0.4em}
\caption{Comparison of the SOTA methods that protect the DNN architecture from memory leakage SCAS attacks, explained in Sec.~\ref{Limitations}. NA indicates not applicable.}
\vspace{-0.5em}
\label{tab:comparison}
\resizebox{\linewidth}{!}{%

\renewcommand\arraystretch{1.1}
\begin{tabular}{cccc}
\hline
Defense & \begin{tabular}[c]{@{}c@{}}Low\\ Latency\end{tabular} & \begin{tabular}[c]{@{}c@{}} Thwarts\\ML SCAS\end{tabular} & \begin{tabular}[c]{@{}c@{}} Thwarts\\NeuroUnlock~\cite{NeuroUnlock}\end{tabular} \\ \hline
Hardware-based~\cite{goldreich1996software,shi2011oblivious,karimi2020hardware} & \color{cadmiumgreen}{\cmark} & \red{\xmark} & NA\\ \hline
Memory Traffic Noise~\cite{DeepSniffer} & $\approx10\times$\red{\xmark} & \color{cadmiumgreen}{\cmark} &  \red{\xmark}\\ \hline
DNN Obfuscation~\cite{li2021neurobfuscator} & \color{cadmiumgreen}{\cmark} & \color{cadmiumgreen}{\cmark} & \red{\xmark} \\ \hline
\textbf{Proposed DNN-Alias} & $\approx1.4\times$ \color{cadmiumgreen}{\cmark} & \textbf{\color{cadmiumgreen}{\cmark}} & \textbf{\color{cadmiumgreen}{\cmark}} \\ \hline
\end{tabular}%
}

\end{table}
\subsection{State-of-the-Art (SOTA) and their Limitations}
\label{Limitations}

\noindent\textbf{Overhead:} Oblivious random access memory (ORAM) schemes encrypt and shuffle the memory read/writes, reducing memory leakage~\cite{goldreich1996software}. However, ORAM-protected designs suffer from high overhead, e.g., $\approx10\times$ latency cost~\cite{liu2015ghostrider}.

\noindent\textbf{Ineffectiveness:} Introducing noise to the execution traces thwarts statistical SCAS attacks, but fails to mitigate ML-based SCAS attacks.\footnote{In first, the attacker applies statistical methods, i.e., correlation analysis, to distinguish the correct secret value among the hypotheses~\cite{Power}, while in second, the attacker trains an ML model to classify the traces.} 
Such ML-based attacks are trained to be resilient to noisy data. For example, the success rate of the ML-based \textit{DeepSniffer}~\cite{DeepSniffer} attack remains the same even with 30\% of amplitude noise.

\noindent\textbf{Limited and Hardware-Specific Security:} 
Current DNN obfuscation methods offer limited security. It has been shown that advanced attacks, such as \textit{NeuroUnlock}~\cite{NeuroUnlock}, can learn the obfuscation procedure and automatically revert it, thereby recovering the original DNN architecture. Further, existing obfuscation techniques, such as \textit{NeurObfuscator}~\cite{li2021neurobfuscator}, depend on extensive hardware profiling. This profiling step makes the defense mechanism dependent on the underlying hardware.

\subsection{Key Research Challenges Targeted in this Work}
The above discussion shows that there is still a gap in designing secure DNN architectures. \textit{Developing an efficient and cost-effective defense mechanism imposes the following important research questions and challenges.}

\begin{enumerate}
 \item \textbf{What makes SCAS attacks successful?} To design secure DNNs, we need to identify the conditions that make the SCAS attacks successful and eliminate them during the development stage of the DNN.

 \item \textbf{Generic security metric.} Defense mechanisms that focus solely on reducing an attack-specific metric, such as the layer error rate (LER),\footnote{The editing distance between extracted and original layer sequence.} cannot mitigate further attack vectors. Thus, devising a generic security metric is required to evaluate the security of DNNs at design time.

 \item \textbf{Performance overhead.} 
A generic and adaptive defense mechanism is required, which can be tailored per target DNN, hardware implementation, and overhead.
 \end{enumerate}

 \subsection{Our Novel Concept and Contributions}

 To address the above challenges, we propose \textit{DNN-Alias}, a DNN obfuscation methodology to protect the architecture of DNNs against static and ML-based SCAS attacks. We argue that SCAS attacks are successful because each layer has a unique run-time trace signature (see Fig.~\ref{fig:Case_Study}). Therefore, we demonstrate that if the signatures of the different layers overlap, it will be difficult for any SCAS attack to differentiate between them. Deterministic DNN obfuscation alters the run-time traces of the layers. However, it does not guarantee overlapping signatures. Our proposed DNN-Alias employs, for the first time, a generic security metric, which measures the overlap between layer signatures by computing the standard deviation. DNN-Alias performs various DNN obfuscation operations to minimize the standard deviation, i.e., resulting in a more secure DNN architecture. Note that the security objective does not depend on any specific attack output but rather on the features of the DNN itself. Furthermore, DNN-Alias does not profile the underlying hardware. Hence, DNN-Alias is a generic defense methodology that is applicable to any DNN running on any hardware. Our novel contributions are summarized in Fig.~\ref{fig:Contribution} and discussed below.

 \begin{enumerate}

\begin{figure}[!t]
 \centering
 \includegraphics[width=\linewidth]{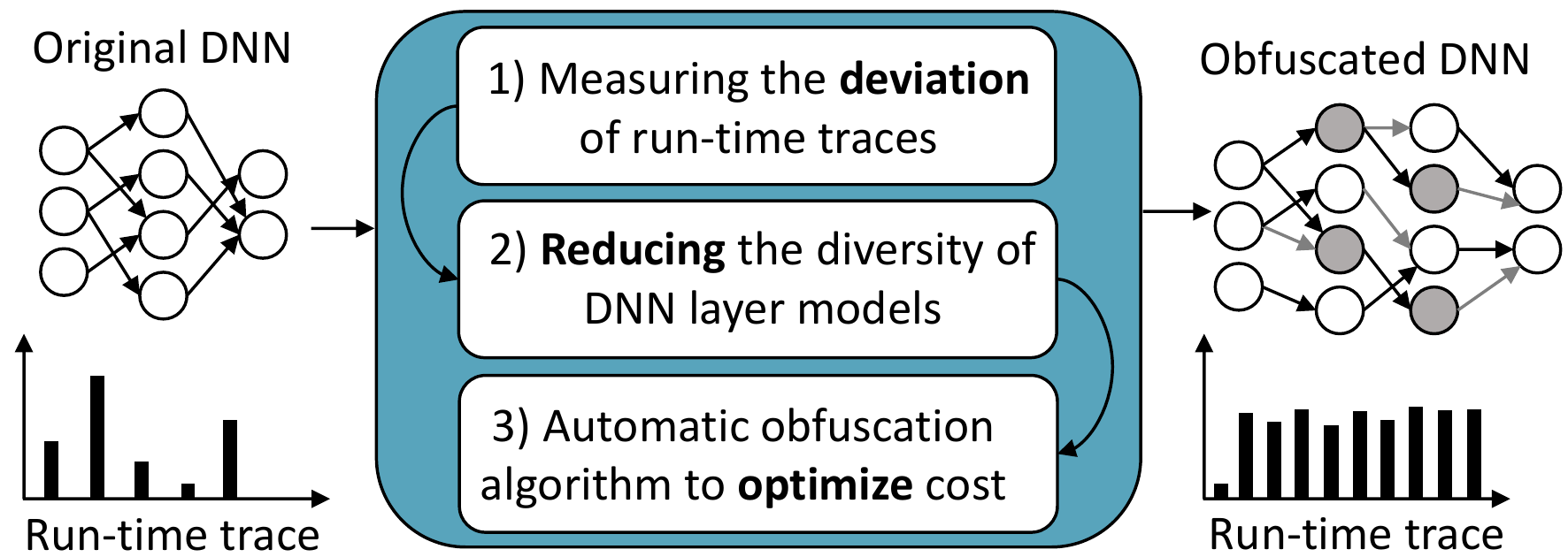}
 \vspace{-1.2em}
\caption{Our contributions presented in this work are in blue box.}
 \vspace{-1em}
\label{fig:Contribution}
\end{figure}

 \item \textbf{General solution to measure the diversity of DNN layers (Sec.~\ref{measuring}):} DNN-Alias defines a novel security metric based on the distribution of run-time parameters to measure the overlap between layer signatures.  Specifically, DNN-Alias calculates the standard deviation of features in the run-time trace and reduces it. By utilizing this metric, DNN-Alias becomes independent of the outcome of any particular attack.

 \item \textbf{Balancing the execution trace (Sec.~\ref{sec:DNN_Obfuscation}):} DNN-Alias presents a novel DNN obfuscation technique that performs layer balancing, limiting information leakage. 
 
 \item \textbf{Efficient DNN obfuscation (Sec.~\ref{GA}):} DNN-Alias employs a genetic algorithm to find an effective combination of obfuscation operations. 
The reward function of this algorithm minimizes the standard deviation in run-time trace while maintaining cost budget.
 \end{enumerate}

\textbf{Key Results:}
We have comprehensively evaluated the efficacy of DNN-Alias with a broad set of random and standard DNNs on image classification, running 
on the Nvidia RTX 2080 TI GPU. We demonstrate that DNN-Alias
increases the LER (between the original and obfuscated DNN) by $2.5\times$ compared to SOTA techniques, resulting in higher obfuscation.

We evaluate the security of DNN-Alias by launching; (i)~an ML-based SCAS attack and (ii)~NeuroUnlock on the obfuscated DNNs. We measure the difference between the architecture of the original DNN and the recovered DNN by the attacks. The LER obtained on the DNN-Alias networks is 2 on average (an LER higher than 1 is considered secure). Further, NeuroUnlock fails to de-obfuscate the networks, reporting an LER of 1.1 on DNN-Alias networks.

Further, DNN-Alias preserves the training accuracy while protecting the DNN against gray-box adversarial attacks. 

%% file: Text/02_Background.tex
\section{Background}
\label{Background}
In this section, we provide the necessary background required to protect DNNs against SCAS attacks.

\subsection{Side-channel-based Architecture Stealing (SCAS) Attacks}
\label{Sec:DeepSniffer}

\begin{figure}[!t]
 \centering
 \includegraphics[width=\linewidth]{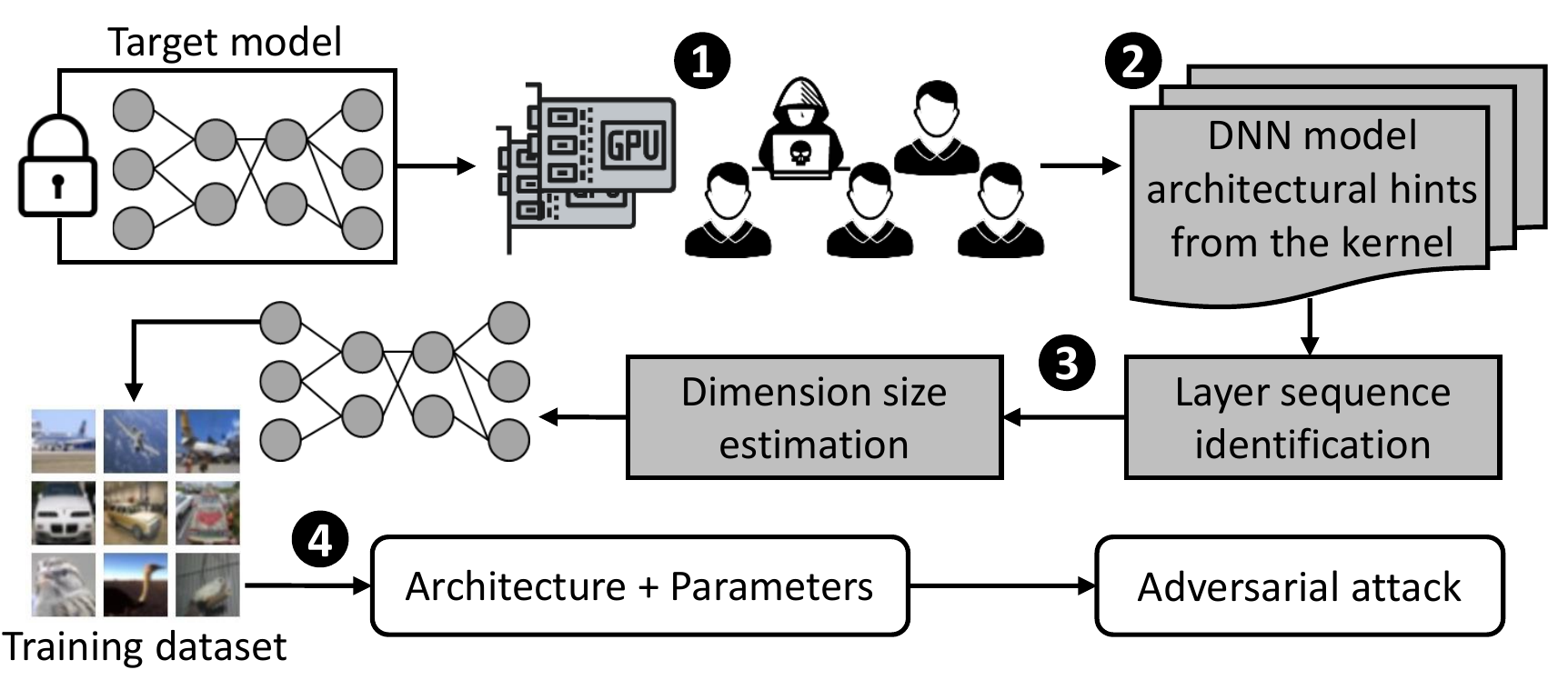}
 \vspace{-2em}
 \caption{Side-channel-based architecture stealing (SCAS) attack. 
}
 \vspace{-0.5em}
 \label{fig:ThreatModel}
\end{figure}

Adversaries can compromise the security of a DNN system by uncovering its confidential model components, such as its architecture and parameters. Extracting an exact copy of the DNN is challenging in a black-box setup, where access to the victim model is limited~\cite{jagielski2020high}. However, physical access to the DNN's hardware platform can lead to the exposure of confidential information through side-channel attacks, such as power analysis or timing analysis~\cite{rakin2021deepsteal} (see~\Circled{\scriptsize\textbf{1}} in Fig.~\ref{fig:ThreatModel}).

Through a SCAS attack, the adversary builds a substitute model using the extracted information and trains it by querying the victim DNN or using a publicly available labeled dataset (see~\Circled{\scriptsize\textbf{2}} and \Circled{\scriptsize\textbf{3}} in Fig.~\ref{fig:ThreatModel}). The attacker can then use the substitute model to launch adversarial attacks against the original DNN system (see~\Circled{\scriptsize\textbf{4}} in Fig.~\ref{fig:ThreatModel}).

\begin{figure}[!t]
 \centering
 \includegraphics[width=0.9\linewidth]{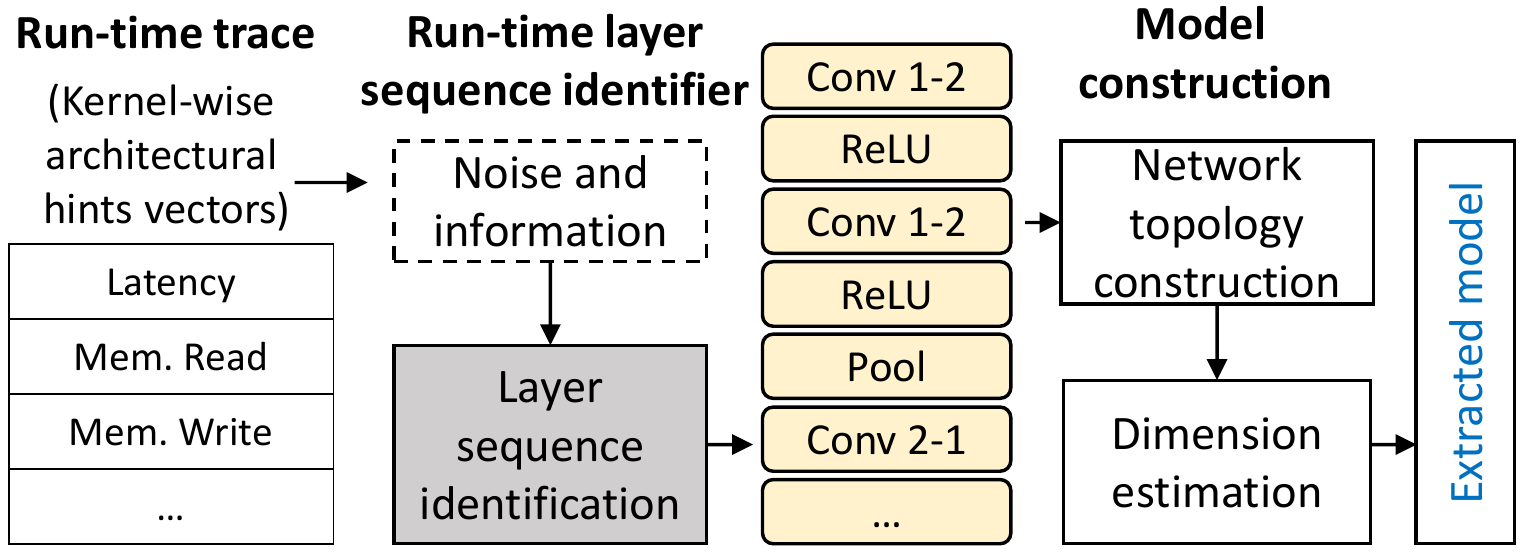}
 \vspace{-0.25em}
 \caption{Flow of the DeepSniffer attack~\cite{deeps}.}
 \vspace{-0.5em}
 \label{fig:DeepSniffer}
\end{figure}

In this study, we focus on attacks that exploit memory, cache, and timing-based information leaks to reveal the architecture of DNNs running on GPU devices. For example, 
 
the DeepSniffer attack~\cite{DeepSniffer}, depicted in Fig.~\ref{fig:DeepSniffer}, employs a long-short-term-memory (LSTM) model~\cite{lstm} to deduce the layer arrangement of a targeted DNN based on its runtime trace.

The run-time trace is a time-series collection of various characteristics, such as execution time and dynamic random access memory (DRAM) access time. The training process for the LSTM model involves profiling randomly generated DNNs on the target GPU. The layer sequence of each generated DNN is encoded as a vector and its runtime trace is extracted. Subsequently, a runtime profile dataset is constructed based on kernel-aware architectural hint, and used to train the LSTM. Attacker utilizes this LSTM as a predictor model, to detect the layer sequence of the DNN from run-time traces. Once the attacker obtains the layer sequence from predictions and construct the model, the dimensions of each layer are determined based on the predicted operation and its position in the time-step and final model is extracted.

To evaluate the accuracy of the SCAS attack, the LER metric has been adopted in literature to measure the difference between the original and extracted DNN layer sequence, which we also use in our analysis. LER is calculated as follows:
\begin{equation}
LER=\frac{ED(L,L^{*})}{|L^{*}|}
\end{equation}
where $L$ represents the predicted layer sequence, $L^{*}$ represents the ground-truth, and $|.|$ denotes the length of a sequence. $ED(p, q)$ denotes the edit distance between the $p$ and $q$ sequences, i.e., the minimum number of insertions, substitutions and deletions required to change $p$ into $q$ (also referred to as the \textit{Levenshtein} distance~\cite{LER}).

%% file: Text/04_Methodology.tex
\section{Our Proposed Defense Mechanism: DNN-Alias}
\label{Methodology}

\begin{figure}[!t]
 \centering
 \includegraphics[width=\linewidth]{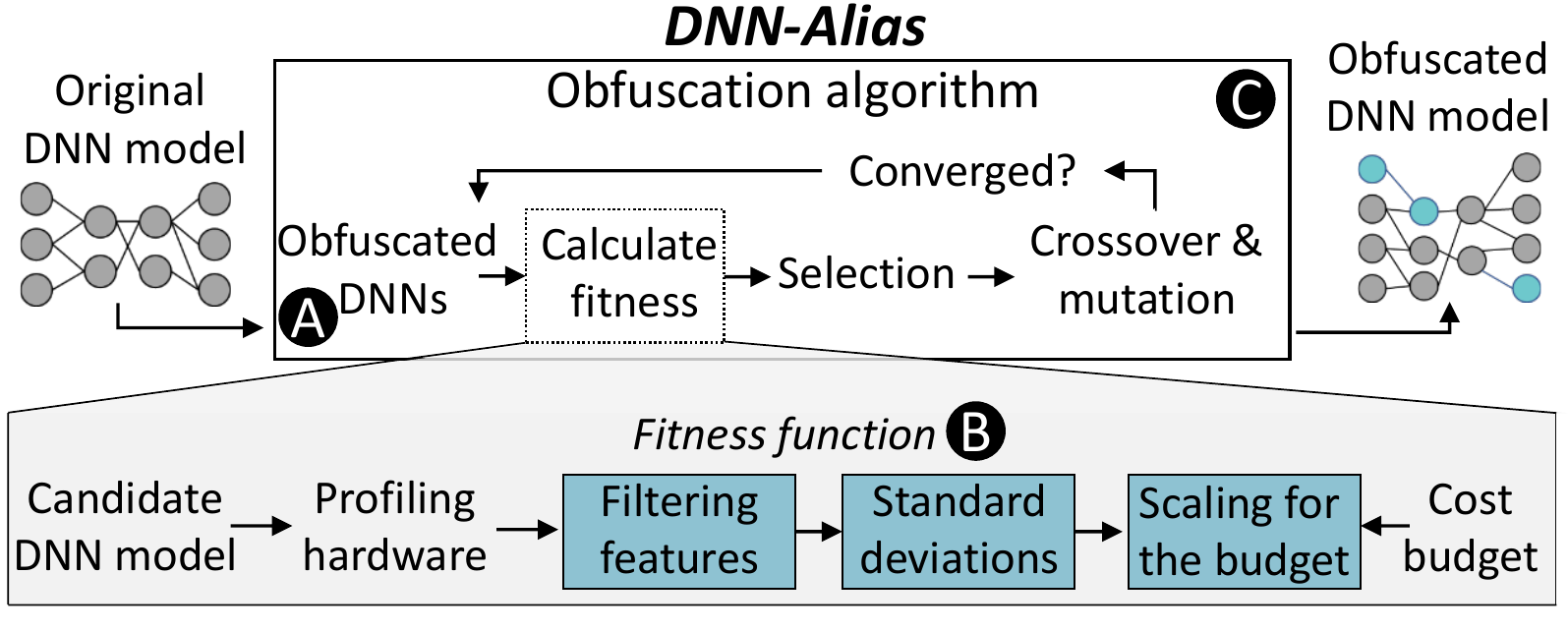}
  \vspace{-2em}
 \caption{Proposed DNN-Alias methodology for DNN architecture protection.} 
  \vspace{-1em}
 \label{fig:Methodology}
\end{figure}

We propose a DNN obfuscation method, DNN-Alias, to thwart SCAS attacks. In this section, we explain the steps of DNN-Alias in detail and summarize them in Fig.~\ref{fig:Methodology}. Further, we discuss the attack model and its assumptions in Sec.~\ref{Adversary-Model}.

DNN-Alias takes an unprotected DNN as input and modifies its architecture by applying layer obfuscation techniques (Sec.~\ref{sec:DNN_Obfuscation}). The goal is to reduce the diversity in the run-time behavior of layers in the given DNN. To measure the diversity of the DNN's run-time behavior, DNN-Alias automatically analyzes its profile during one inference execution (Sec.~\ref{measuring}). DNN-Alias employs a genetic algorithm to guide the obfuscation and balance the DNN layers, taking into account the overhead budget (Sec.~\ref{GA}).

\subsection{Threat Model and Assumptions}
\label{Adversary-Model}

Consistent with most recent related works~\cite{dataset}, we assume that the adversary has no prior knowledge of the victim DNN architecture, parameters, training algorithms, or hyper-parameters. We focus on edge security, in which the attacker has (i)~system privilege access to the GPU platform encapsulating the victim DNN, (ii)~the inputs and outputs (labels) of the DNN, and (iii)~a publicly available training dataset. \textit{We show that even with such a powerful threat model, attackers cannot steal the architecture of the DNNs obfuscated via our proposed DNN-Alias.}

\subsection{DNN Layer Sequence Obfuscation Operations}
\label{sec:DNN_Obfuscation}

DNN-Alias uses function-preserving obfuscation operations to protect the architecture of DNNs. DNN-Alias carefully apply one (or more) operation to each layer of the original model, making the target DNN more difficult to reverse engineer or attack (Step~\Circled{\scriptsize\textbf{A}} in Fig.~\ref{fig:Methodology}).

Please note that DNN obfuscation operations have been used before to protect the DNN architecture~\cite{zhu2019eena,li2021neurobfuscator}. DNN-Alias applies the same obfuscation knobs to change the run-time profile of each layer function and hide the architecture of original DNN. However, existing solutions either randomly obfuscate the layers or focus on a specific SCAS attack, resulting in weak protection. DNN-Alias guides the obfuscation differently, thwarting all SCAS attacks. Next, we explain the obfuscation operations used by DNN-Alias. 

\begin{figure}[!t]
 \centering
 \includegraphics[width=\linewidth]{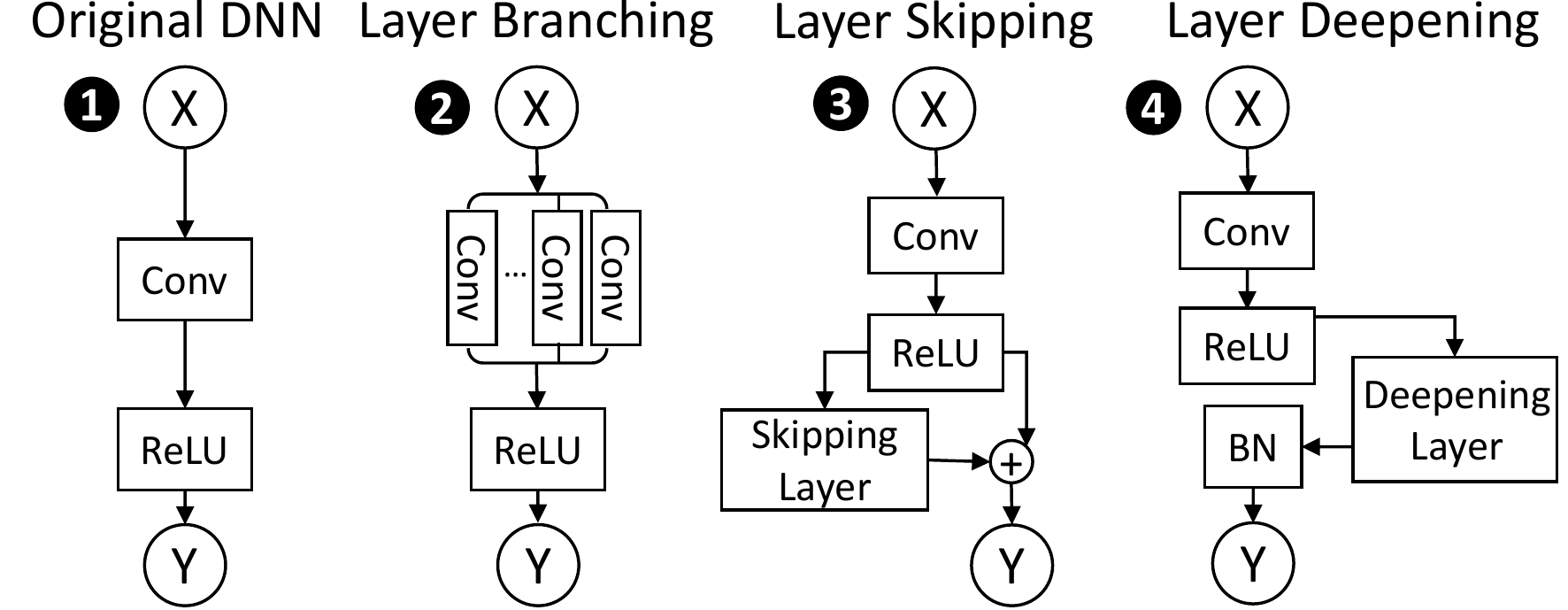}
 \vspace{-1.7em}
 \caption{Visualization of the employed obfuscation operations~\cite{zhu2019eena, li2021neurobfuscator}.}
 \vspace{-1.2em}
 \label{fig:Obf_Operation}
\end{figure}

Let the matrix $\mW_{k1, k2, c, j}^{(i)}$ represent the $i^{th}$ convolutional layer to be modified. $k1$ and $k2$ represent the height and width of the convolution kernel, respectively, while $c$ and $j$ denote the input and output channel size, respectively. $\mX^{(i)}$ and $\sigma(\cdot)$ denote the input of the layer and the activation function (e.g., \textit{ReLU}), respectively. Fig.~\ref{fig:Obf_Operation}~\Circled{\scriptsize\textbf{1}} illustrates the original operator.

\textbf{Layer Branching.} dividing a single layer operator into smaller, partial operators, as demonstrated in Fig.~\ref{fig:Obf_Operation}~\Circled{\scriptsize\textbf{2}}. For example, a 2-D convolution layer (\textit{Conv2D}) $\mW_{k_1, k_2, c, j}^{(i)}$  can be separated into two partial convolutions, as follows.
\begin{equation}
\begin{split}
\mU_{k_{1},k_{2},c,j/2}^{(i)} = \mW_{k_{1},k_{2},c,m}^{(i)} \quad m \in \left[0,\lfloor \frac{j}{2} \rfloor \right),\\
\mV_{k_{1},k_{2},c,j/2}^{(i)} = \mW_{k_{1},k_{2},c,m}^{(i)} \quad m \in \left[\lfloor \frac{j}{2} \rfloor,j \right)
\end{split}
\end{equation}
The final output is obtained by combining the two partial results.
\begin{equation}
 \mU_{k_1, k_2, c, j/2}^{(i)} * \mX^{(i)} || \mV_{k_1, k_2, c, j/2}^{(i)} * \mX^{(i)}
\end{equation}
The splitting can also be performed in the input channel dimension. 
The final output in this case is the addition of the two, as follows, where $\mX^{(i)}$ is sliced into $\mA^{(i)}$ and $\mB^{(i)}$.
\begin{equation}
 \mU_{k_1, k_2, c/2, j}^{(i)} * \mA^{(i)} + \mV_{k_1, k_2, c/2, j}^{(i)} * \mB^{(i)}
\end{equation}

\textbf{Layer Skipping.} 
An additional Conv2D layer $\mU^{(i+1)}_{k_1, k_2, j, j}$ with all its parameters set to $0$ is inserted to retain the original functionality. An illustration of this is shown in Fig.~\ref{fig:Obf_Operation}~\Circled{\scriptsize\textbf{3}}. The Conv2D layer can be expressed as follows, with $\sigma \left( \mX^{(i+1)} \right)$ representing the activation output of $i^{th}$ original layer.

\begin{equation}
\sigma \left( \mX^{(i+1)} \right) + \sigma \left( \mU^{(i+1)} * \mX^{(i+1)} \right) = \sigma \left( \mX^{(i+1)} \right) 
\end{equation} 

\textbf{Layer Deepening.} adds a new computational layer to the sequence.
 The new layer is inserted after the activation of the current layer and before the \textit{batch normalization} (BN) step as shown in Fig.~\ref{fig:Obf_Operation}~\Circled{\scriptsize\textbf{4}}. If the previous layer is linear, the newly added layer $\mU^{(i+1)}$ is initialized as an identity matrix $\mI$ to preserve the function of the model. Otherwise, $\mU^{(i+1)}_{k_1, k_2, j, j}$ can be generalized as:
\begin{equation}
\mU_{a,b,c,d}^{(i+1)}=
\left\{ 
 \begin{array}{lr}
 1 & a= \frac{k_{1}+1}{2} \wedge b= \frac{k_{2}+1}{2} \wedge c=d \\
 0 & \rm otherwise
 \end{array}
\right. 
\end{equation}
Layer deepening is effective as long as the activation function satisfies the following condition, like the ReLU function.

\begin{equation}
\forall x:\sigma(x)=\sigma \left( \mI*\sigma(x) \right)
\end{equation}

Post obfuscation, the computation graph is extracted and fed to the TVM\textsuperscript{\textregistered}~compiler~\cite{chen2018tvm}. The compiler performs optimizations at the graph and operator levels, generating low-level optimized code for GPU execution.\footnote{DNN obfuscation can be categorized into sequence and dimension obfuscation. We focus on sequence obfuscation since the sequence identification stage is the most fundamental step in SCAS attacks.}

\subsection{Measuring the difference of the layers}
\label{measuring}

\begin{figure}[!t]
 \centering
 \includegraphics[width=\linewidth]{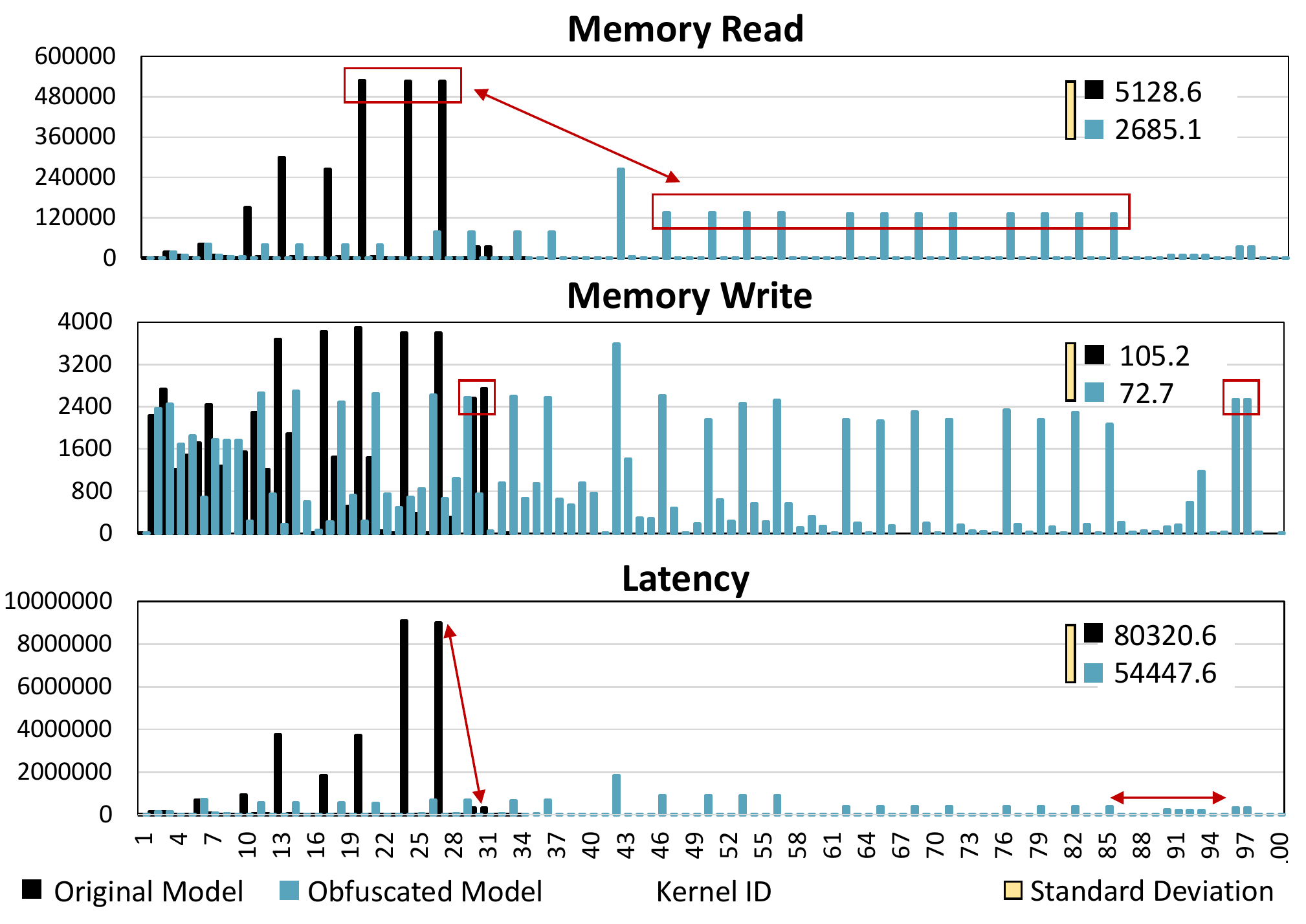}
 \caption{Comparison of the run-time trace of each kernel for memory access time and computational latency. The DNN reads the inputs and weights of each layer from the memory (Memory Read), executes the function of the layer (\#Cycles) and writes the output of the layer in memory (Memory Write). DNN-Alias decreases the standard deviation in each trace.  
 }
 \label{fig:Traces}
\end{figure} 

DNN-Alias guides the obfuscation algorithm to assemble a configuration of obfuscation knobs on a given DNN (Sec.~\ref{GA}). For each obfuscated DNN, DNN-Alias analyzes its run-time execution trace in inference mode and computes the difference of the levels in the trace using a generic measurement technique (See ~\Circled{\scriptsize\textbf{B}} in Fig.~\ref{fig:Methodology}).
In this technique, the run-time traces are collected and analyzed online and target hardware is not required to be profiled in advance.

DNN-Alias measures the difference of the levels in a run-time trace using the standard deviation ($St.D$) of values in a kernel per each feature in the run-time profile of the DNN.\footnote{A standard deviation is a measure of how dispersed the data is in relation to the mean. A low standard deviation means data are clustered around the mean, and a high standard deviation indicates data are more spread out.} 

\begin{equation}
\begin{aligned}
St.D = \sqrt{\frac{1}{N-1} \sum_{i=1}^N (x_i - \overline{x})^2} 
\end{aligned}
\label{eq:St.D}
\end{equation}

$N$ is number of kernels, $x_i$ is the value in run-time trace and $\overline{x}$ is the mean of the values. First, the original DNN is configured by incorporating the obfuscation operations. Then, the obfuscated DNN is executed on the hardware platform and $St.D$ of the above features is calculated. DNN-Alias uses a genetic algorithm to find the best configuration of the obfuscation operations, by decreasing the $St.D$ value so that the run-time traces for all layers are similar. Decreasing the $St.D$ (difference) enables layer balancing and hides the function of the kernel from the attacker's predictors.

In Fig.~\ref{fig:Traces}, we present an example of DNN-Alias obfuscation by comparing the execution trace of the unprotected (black) and obfuscated (blue) ResNet-20. The execution of each layer in the network involves reading weights and inputs from memory through a kernel process, executing the layer's function, and writing the output back to memory. 
The combination of these three values for each trace per kernel creates a unique pattern, which reveals information to SCAS attacks and allow them to predict the function of the layer.
To counter this, DNN-Alias reduces the difference between the traces and forces all layers to have similar run-time traces.

Further, in Fig.~\ref{fig:Traces} we observe that in the trace of unprotected DNN (shown in black), initially there are high values (related to Convolution functions) that are noticeably distinct from the two final (functions related to the fully connected) layers (red arrow in Latency trace). However, in the trace of obfuscated DNN (shown in blue), the processing of large inputs is split into smaller functions (red box in Memory Read trace) while some functions are not changed (red boxes in Memory Write trace). This technique, reduces the differences ($St.D$ values shown by yellow box) and makes these two functions almost indistinguishable in the traces. For example, the $St.D$ value of the latency trace drops from $80320.6$ to $54447.6$, i.e., the $St.D$ was reduced by $32\%$, thwarting SCAS attacks.

\subsection{Efficient exploration of obfuscations}
\label{GA}
 
The objective of DNN-Alias is to identify the optimal arrangement of obfuscation functions while staying within the cost limit. The cost taken into account is the inference latency of the DNN. This optimization challenge can be framed as a limited discrete optimization problem that minimizes the $St.D$ of the layers with the termination condition of cost budget, described below. $N$ is the number of features considered in profilers, $S$ is the set of obfuscation knobs,
$T$ is the latency with obfuscation, and $T^*$ is the latency of original DNN. 

\begin{equation}
\begin{aligned}
\min_{S} \quad & \sum_{i=1}^{N}{St.D_{i}(S)}\\
\textrm{s.t.} \quad & T\leq (1 + B) T^*\\
\end{aligned}
\label{eq:fitness1}
\end{equation}

\input{Figures/algorithm}

To achieve this, we utilize a genetic algorithm to search the space of potential solutions for random combinations of obfuscation functions and determine the difference in the layer kernel-wise values (as outlined in Algorithm~\ref{algorithm:genetic_algorithm}).

\textbf{Initial population.} For a given DNN, first, DNN-Alias automatically creates a random starting population of obfuscated DNNs with 16 candidates, i.e., population size $size_{p}=16$, all with the same original DNN functionality. The initial random obfuscation procedure is explained next.

\textbf{Random DNN obfuscation.} A random selection process is utilized to determine the insertion and configuration of obfuscation operations within each layer of the specified DNN. For each obfuscation operation, a binary random decision is made with a $50\%$ likelihood to determine if the operation will be utilized or disregarded. As a result, a layer may incur the application of 1 and 2 (each $37.5\%$ probability), 3 or none (each $12.5\%$ probability) obfuscation operations.

\textbf{Fitness function.} Then, we evaluate each member of the population using the fitness score, which is determined by the $St.D$ and a scaled value of the cost budget, as defined below.

\begin{equation}
\displaystyle Fitness =\quad\sum_{i=1}^{N}{St.D_{i}(S)} . \biggl[ \biggl(\dfrac{T - (1 + B) T^* }{T^*}\biggr) ^2 \biggr]
\label{eq:fitness2}
\end{equation}

\begin{figure}[!t]
 \centering
 \includegraphics[width=\linewidth]{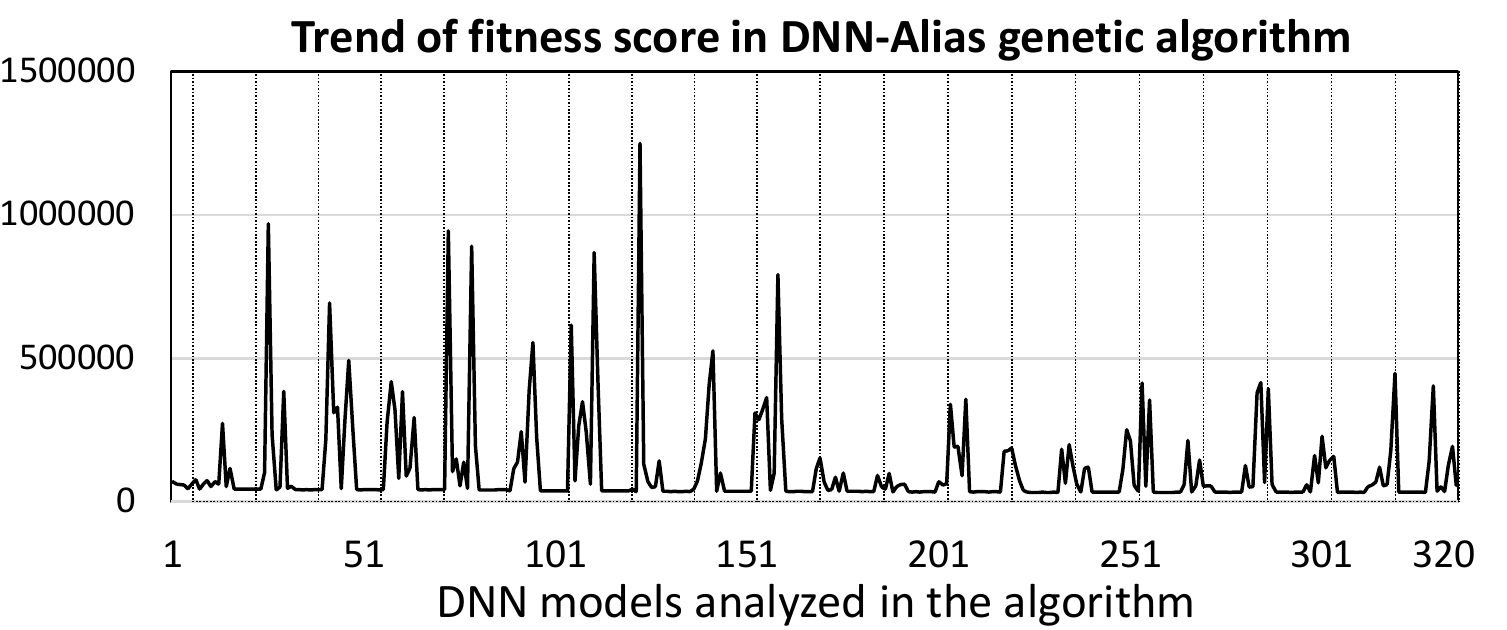}
\caption{Fitness score trend of DNN-Alias on ResNet-20 with a population of 16. Here, we have 320 elements and 20 mutations.}
 \vspace{-1em}
\label{fig:LER_Real}
\end{figure}

\textbf{Crossover and mutation.} For the mating process, we rank the population based on their fitness scores and select the top half of individuals with the best scores as parents and add them to the next candidate pool. Also, parents are brought to the crossover process, where they combine to form an equal number of offspring that are added to the pool of candidates. The offspring are built from the crossover of the parents' obfuscation list, then mutated by adding Gaussian noise. These methods are known as 1-point crossover and Gaussian mutation in literature. 
As shown in Fig.~\ref{fig:LER_Real}, the trend of the fitness score gradually decreases with each pool, and the best candidates are carried forward to the next generation. 

\textbf{Final solution.} The mutation process continues until the fitness score converges and stabilizes, which we found to occur after 20 generations in our case.

%% file: Figures/algorithm.tex
\begin{figure}[!t]
    \begin{algorithm}[H]
        \begin{footnotesize}
        \caption{DNN-Alias Genetic Algorithm-Based Obfuscation }
        \label{algorithm:genetic_algorithm}
        \begin{algorithmic}[1]
            \Procedure{Obfuscation}{$DNN$, $budget$, $size_{p}$, $generations$}
                \State $population$ $\gets$ generate $size_{p}$ variations of $DNN$
                \For{$generation_{index}$ in $0...generations$}
                    \State $list_{rewards}$ $\gets$ empty list
                    \For{$population_{element}$ in $population$}
                        \LComment{\blue{profiling is done with nvprof on GPU}}
                        \State $prof$ $\gets$ $profile(population_{element})$
                        \State $reward_{layer balancing}$ $\gets$ sum standard deviations in $prof$
                        \State $scaling_{budget}$ $\gets$ $scaling(latency, budget)$
                        \State $reward$ $\gets$ $reward_{layer balancing} \cdot scaling_{budget}$
                        \State $list_{rewards}$ append $reward$
                    \EndFor
                    \LComment{\blue{population update}}
                    \State $population$ $\gets$ select top 50\% based on $list_{rewards}$
                    \State $population$ $\gets$ crossover and mutate $population$
                \EndFor
            \EndProcedure
        \end{algorithmic}
        \end{footnotesize}
    \end{algorithm}
    \vspace{-1.5em}
\end{figure}

%% file: Text/05_ExperimentalSetup.tex
\section{Experimental Setup}
\label{Setup}

In this section, we present our experimental setup in evaluating the effectiveness and security of DNN-Alias.

\textbf{Hardware.} For our experiments, we utilized the Nvidia RTX 2080 Ti GPU as our experimental platform. However, our proposed method is generic and applicable to other GPUs and hardware platforms. We employ Nsight\textsuperscript{\textregistered}~Compute~\cite{Nsight} for profiling the GPU and launching the ML-based SCAS attack, which requires privileged access to the performance counters. A dataset of randomly generated DNNs was generated to profile them on the GPU and train the attack model predictors. Next, we explain how we create the dataset of random DNNs.

\textbf{Obfuscation Algorithm.}
\label{Randoms}
We implement DNN-Alias using the \textit{PyTorch} deep learning framework~\cite{pytorch}. The DNNs (original and obfuscated) are described as Python model files. The genetic algorithm is implemented by using \textit{pymoo}~\cite{pymoo}.

\textbf{Random DNN Creation.} To train the predictors of SCAS attack, we use the following method for generating 5,000 different DNNs for image classification, using the CIFAR-10 dataset as a reference. The number of Conv2D layers in each DNN is randomly selected from the range [4, 12], and the number of fully connected (FC) layers is randomly selected from the range [1, 4]. The output channel sizes of Conv2D layers and the dimensions of FC layers are also randomly chosen from a range of predefined values. Some Conv2D layers are randomly replaced with blocks from the ResNet and MobileNet networks, and some Conv2D layers are changed to pooling layers. Batch normalization (BN) layers are added after each Conv2D and FC layer. All DNNs have 3 input channels, width and height of 32, and 10 output classes. Using this technique, we also generate 700 random DNNs to analyze the performance of DNN-Alias in obfuscating random DNNs.

\textbf{SCAS-based Adversarial Attack.}
 \label{Adv}
 In an adversarial attack scenario, the attacker manipulates the output of a DNN by adding subtle, almost imperceptible alterations to the input images. The objective of the attack is to find the smallest possible changes in the input that can cause the DNN to produce incorrect output, either arbitrarily (in case of untargeted attack) or as pre-determined (in case of targeted attack).  To launch an adversarial attack on a DNN that operates as a gray box, the attacker often develops a substitute model by examining the input and output of the victim DNN. With the help of the SCAS attack, the adversary has access to the details of target DNN with high accuracy to build the substitute model.
Then, adversarial samples are created using the white-box substitution technique. Finally, these adversarial samples are utilized to disrupt the workings of the target DNN.

\begin{figure}[!t]
     \centering
     \includegraphics[width=\linewidth]{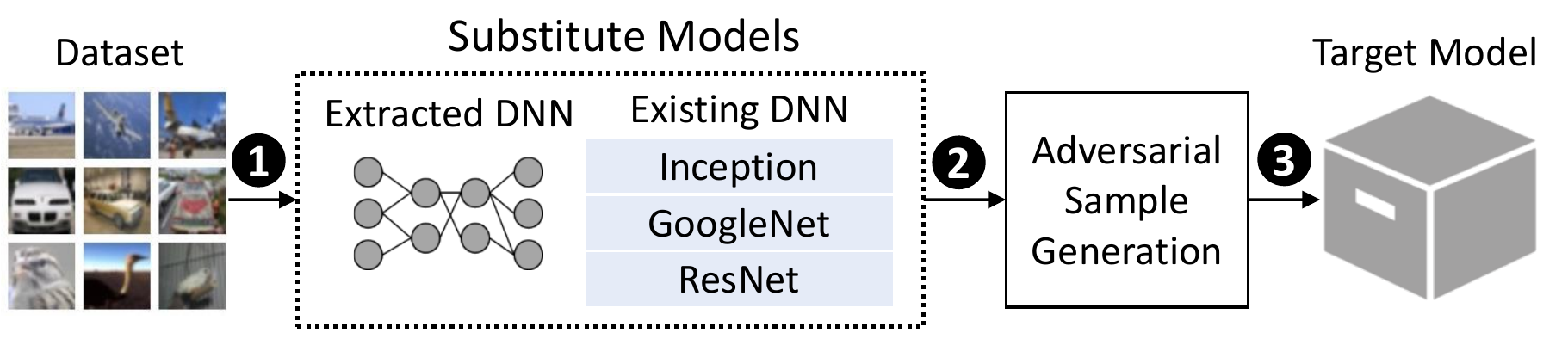}
      \vspace{-1.5em}
     \caption{Adversarial attack flow on the extracted DNN architecture. }
      \vspace{-1em}
     \label{fig:Adversarial}
\end{figure}

In summary, Fig.~\ref{fig:Adversarial} depicts the transfer-based adversarial attack flow, which includes the following steps:

\subsubsection{Substitute Models} In this step, we train substitute models to closely mimic the target model's behavior. For black-box adversarial attacks, the substitute models are selected from publicly available DNN families. In SCAS-based adversarial attacks, the substitute model is obtained from runtime traces. We compare the success rate of adversarial attacks in both scenarios. 

\subsubsection{Adversarial Sample Generation} The most advanced methods utilize an ensemble approach to increase the likelihood of a successful attack, based on the idea that if an adversarial image is able to fool multiple models, it is more likely to have a similar effect on the black-box model. We follow the same procedure to produce adversarial images for our target DNNs.
\subsubsection{Deployment of Adversarial Samples} We pass the generated adversarial examples as input data to launch an attack on the gray-box DNN.

%% file: Text/06_Results.tex
\section{Security Analysis and Overhead Results}
\label{Results}

\begin{figure}[!t]
 \centering
 \includegraphics[width=\linewidth]{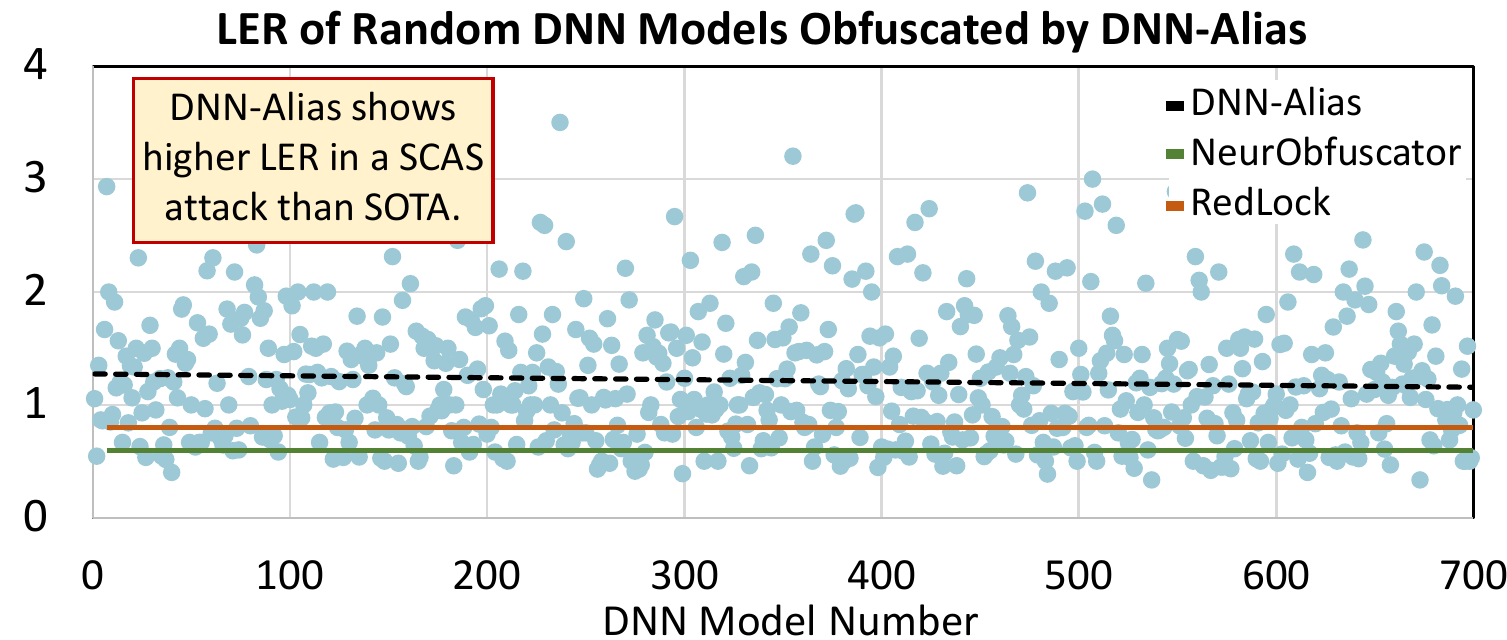}
 \vspace{-2em}
\caption{LER of random DNNs obfuscated by DNN-Alias and SOTA techniques.}
 \vspace{-1em}
\label{fig:Rand_NNs}
\end{figure}

In this section, we present the security evaluation and performance analysis of the DNN-Alias compared to SOTA.

\subsection{Effectiveness of DNN-Alias}
\begin{figure}[!t]
 \centering
 \includegraphics[width=\linewidth]{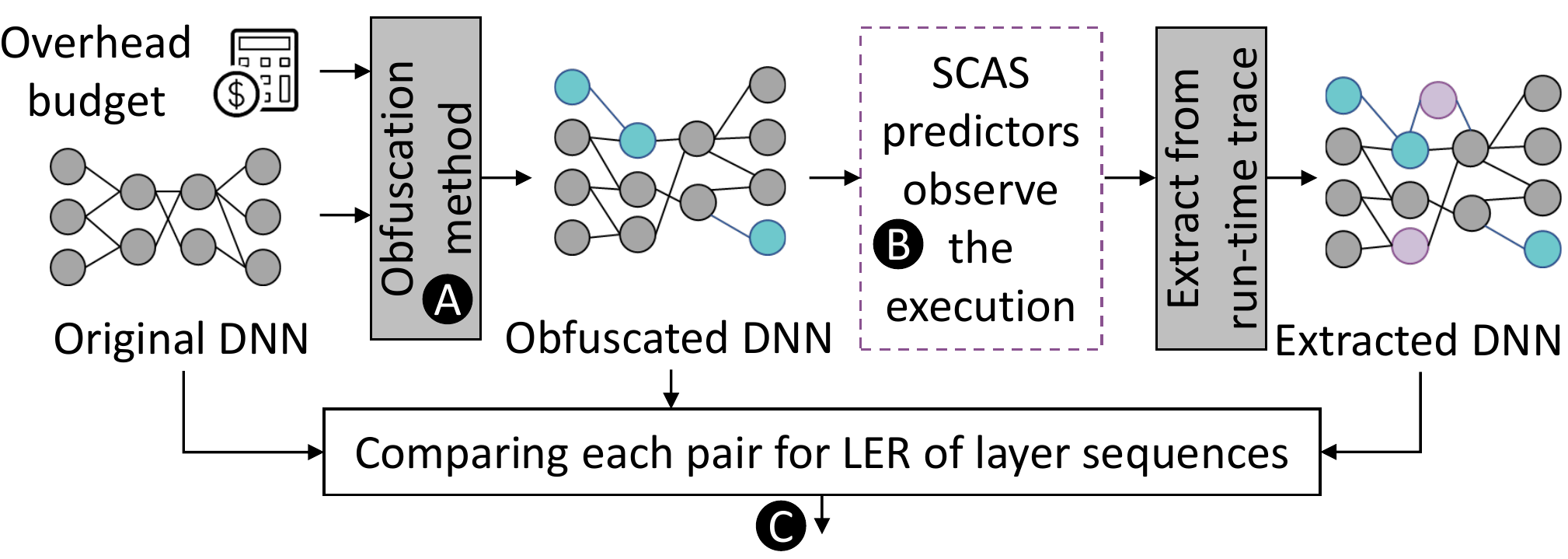}
 \vspace{-1.5em}
 \caption{Measuring the accuracy of the SCAS attack, when the DNN is protected by obfuscation methods.}
 \vspace{-1em}
 \label{fig:LER_Specified}
\end{figure}

In Fig.~\ref{fig:LER_Specified}, we show the obfuscation and SCAS attack evaluation procedure. The original DNN is first obfuscated by DNN-Alias (step~\Circled{\scriptsize\textbf{A}}). Next, the runtime traces are collected to extract the DNN architecture via an ML-based SCAS attack (step~\Circled{\scriptsize\textbf{B}}). 
By comparing the LER between the original and extracted DNNs, we can analyze the effectiveness of the obfuscation method (step~\Circled{\scriptsize\textbf{C}}). The best case scenario for the SCAS attack is to obtain an LER close to $0$.

 \subsubsection{Effectiveness on Random DNNs} 
We obfuscate $700$ randomly generated DNNs (explained in Sec.~\ref{Randoms}) using DNN-Alias, launch the SCAS attack, and measure the security in terms of the LER (Extracted\_Obf, Original). These results are shown in Fig.~\ref{fig:Rand_NNs}. The cost budget considered in this experiment is set to $0.2$. 
The minimum LER value observed in this experiment is $0.3$ and the maximum is $3.5$. The FORECAST calculation of LER (predicts a future value by using linear regression) shows $1.2$. The average of LER for SOTA techniques, is shown in Fig.~\ref{fig:Rand_NNs}, where LER for NeurObfuscator~\cite{li2021neurobfuscator} is $0.62$ and for ReDLock~\cite{NeuroUnlock} is $0.73$. Therefore, \textit{DNN-Alias obfuscation is $\approx2\times$ more resilient against SCAS attacks compared to SOTA.}

\subsubsection{Effectiveness on Publicly Available DNNs}

Further, we analyzed the effectiveness of DNN-Alias on a set of real DNNs as a case study. The results are shown in Fig.~\ref{fig:LER_Compared}.

First, we launch the ML-based SCAS attack on the original DNNs, i.e., without any obfuscation. The red bar for LER (Extracted\_org, Original) shows the ML-based SCAS predictor errors with an average of $0.05$, indicating that the original DNNs are completely vulnerable to SCAS attacks. Our goal is to increase this LER value over $1$.

Next, each DNN was obfuscated using DNN-Alias for two cost budgets ($0.2$ and $0.6$). We launch the same ML-bases SCAS attack on the obfuscated DNNs. 

The blue bar LER (Extracted\_obf, Original) represents the difference between the extracted DNN by the SCAS attack and the original DNN, which averages $1.8$ for DNN-Alias. The increase of the LER from an average of $0.05$ to $1.8$ demonstrates that DNN-Alias is highly effective in protecting the DNN against SCAS.

\subsubsection{Obfuscation Overhead and Cost Budget} In all cases presented in Fig.~\ref{fig:LER_Compared}, we can see that increasing the cost budget (from $0.2$ to $0.6$) increases the LER (from an average of $1.7$ to $1.96$), i.e., leads to stronger obfuscation.

The gray bar LER (Obfuscated, Original) in Fig.~\ref{fig:LER_Compared} represents the difference between the original DNN and the obfuscated DNN. The LER, in this case, has an average of $0.1$, demonstrating that \textit{DNN-Alias effectively thwarts SCAS attacks through minor changes in the original DNN}, avoiding high overhead costs. 

\subsubsection{NeuroUnlock Attack on DNN-Alias}

\begin{figure*}[!t]
 \centering
 \includegraphics[width=\textwidth]{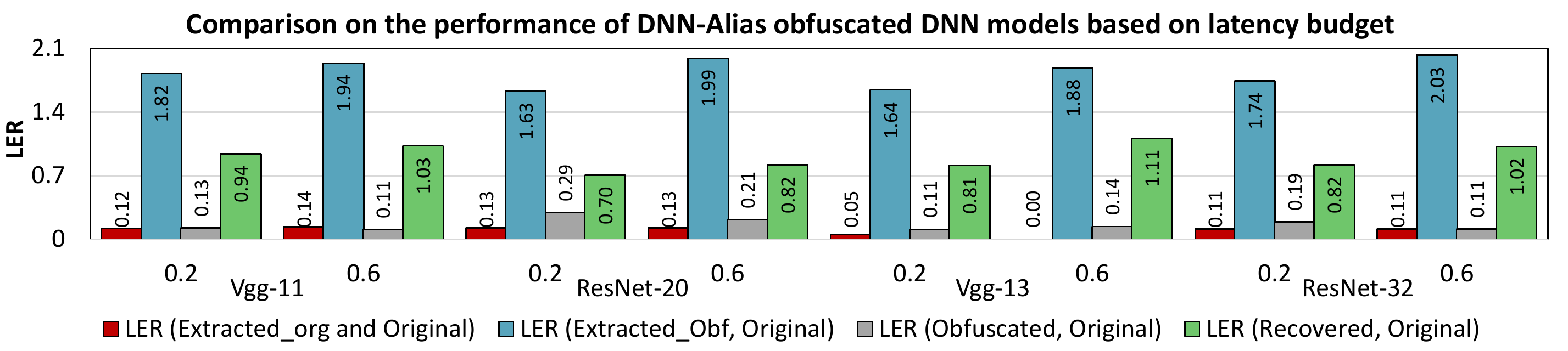}
 \vspace{-1.5em}
\caption{The LER for the DNN-Alias obfuscated DNNs. }

 \vspace{-1em}
\label{fig:LER_Compared}
\end{figure*}

We evaluated the SOTA NeuroUnlock attack~\cite{NeuroUnlock} on DNN-Alias and NeurObfuscator~\cite{li2021neurobfuscator}. NeuroUnlock attempts to reverse the obfuscation of the extracted DNN from the SCAS attack using sophisticated ML-based models. The green bar labeled ``LER (Recovered, Original)'' in Fig.\ref{fig:LER_Compared} shows the difference between the original DNN and the recovered DNN using NeuroUnlock when DNN-Alias is in place. With an average LER of $0.9$, the results indicate that NeuroUnlock failed to accurately recover the original DNN. In comparison, recovering the obfuscated DNN using NeuroUnlock with NeurObfuscator in place resulted in an average LER of $0.31$. This suggests that \textit{DNN-Alias is $\approx3\times$ more robust against de-obfuscation techniques than the current SOTA methods.}

 \subsubsection{Comparison with SOTA}

We compare the effectiveness of DNN-Alias to NeurObfuscator~\cite{li2021neurobfuscator} considering the target real DNNs. We obfuscate the DNNs with $0.2$ and $0.6$ latency budgets using DNN-Alias and NeurObfuscator (the code is open-sourced). We apply the SCAS attack on the obfuscated DNNs and compare each extracted DNN to the original DNN. 

\begin{figure}[!t]
 \centering
 \includegraphics[width=\linewidth]{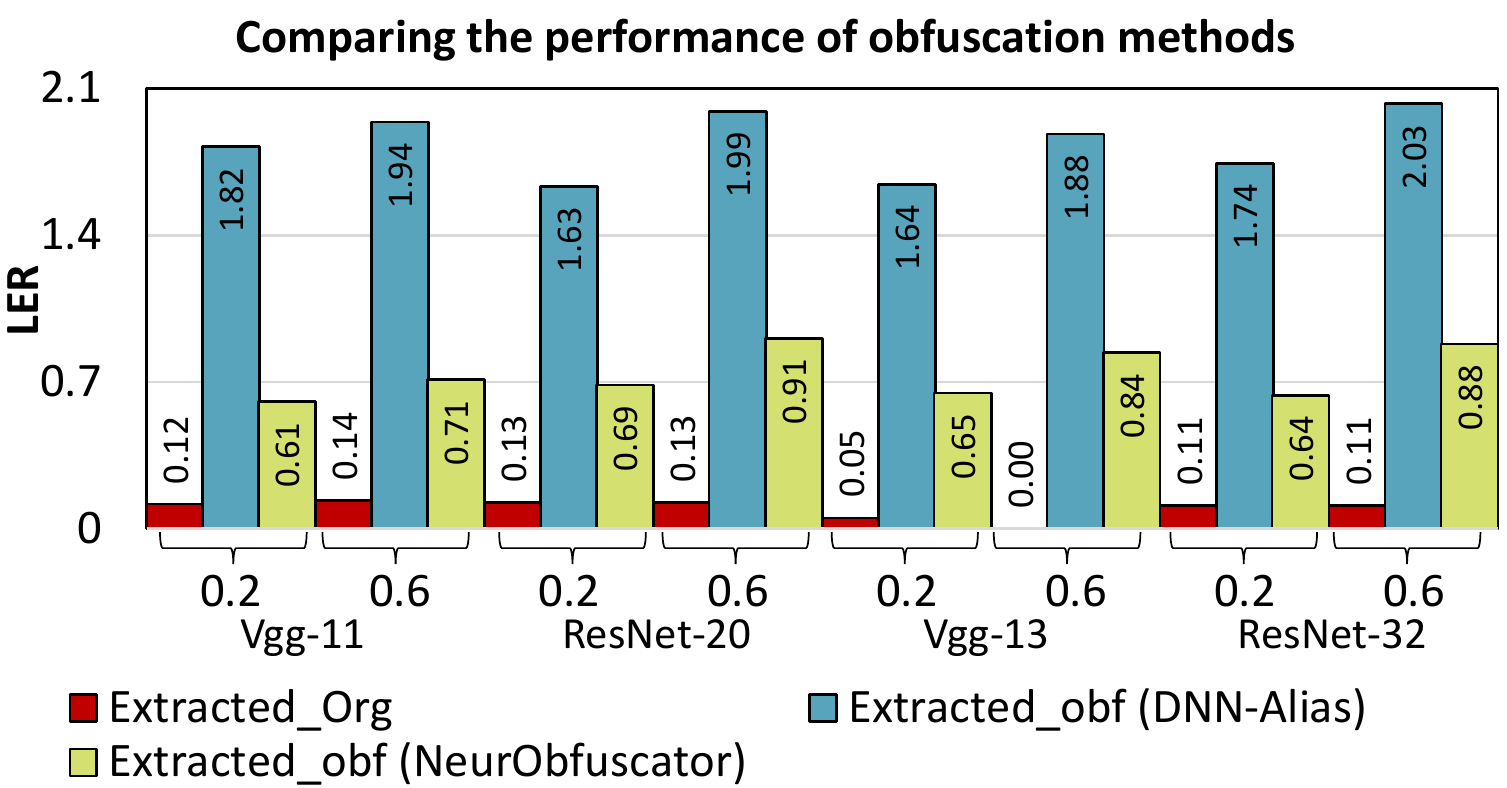}
 \vspace{-1.5em}
\caption{The LER for the DNN-Alias and NeurObfuscator obfuscated DNNs compared to the original model.}
\label{fig:LER_Real_NNAlias}
\vspace{-1.5em}
\end{figure}

The results in Fig.~\ref{fig:LER_Real_NNAlias} show that the LER of obfuscated DNNs using DNN-Alias is $2.5\times$ (on average) more than NeurObfuscator. \textit{In summary, DNN-Alias is more effective in hiding the layer sequence of DNNs compared to SOTA.}

\subsection{Performance Analysis}

In this section, we examine the effect of DNN-Alias on the DNN training and the overhead involved in its design. Last but not least, we study the success rate of adversarial attacks against DNN-Alias networks.

\subsubsection{Training Performance}

\begin{figure}[!t]
 \centering
 \includegraphics[width=\linewidth]{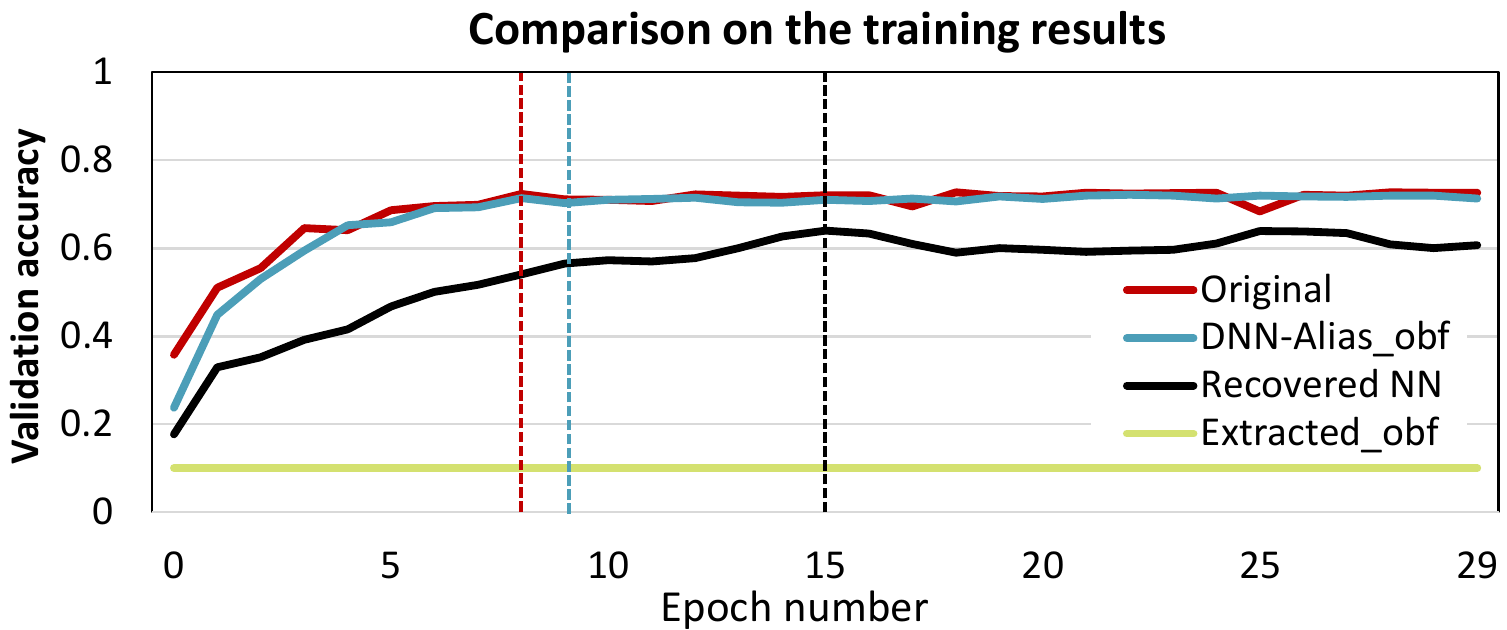}
 \vspace{-2em}
\caption{Validation accuracy after training the models.}
 \vspace{-1em}
\label{fig:Accuracy}
\end{figure}

To assess the effect of DNN-Alias on the functionality of the DNN, we compare the validation accuracy of the original and obfuscated DNNs. Further, we train the DNNs recovered by the SCAS attack and by NeuroUnlock attack to see if the high LER values map to loss in DNN performance. Fig.~\ref{fig:Accuracy} shows the validation accuracy of the VGG-11 DNN over $30$ epochs of training on the CIFAR-10 dataset~\cite{cifar10}. 

The results demonstrate that DNN-Alias (blue line) maintains the functionality of the DNN and does not impact the training performance. However, the recovered DNN by the SCAS attack (green line) does not work and simply does not converge, with a $75\%$ drop in performance. Further, even if NeuroUnlock was launched after the SCAS attack to revert the obfuscation, the recovered model (black line) still shows a lower validation accuracy, with a drop of $10\%$, and converges about $8$ epochs later.
Therefore, \textit{DNN-Alias forces the SCAS attack to recover a DNN with worse performance compared to the original DNN}.

\subsubsection{Adversarial Attack}
We launch the SCAS-based adversarial attacks (discussed in Sec.~\ref{Adv}) on VGG-11 DNN obfuscated by DNN-Alias as target model.

In this experiment, the adversarial samples are generated from the CIFAR10 dataset and the label output of VGG-11. We study this attack on the original (un-protected) DNN and then compare it with the obfuscated DNN. Also, we show the success rate of this attack on similar DNN families. 

\begin{figure}[!t]
 \centering
 \includegraphics[width=\linewidth]{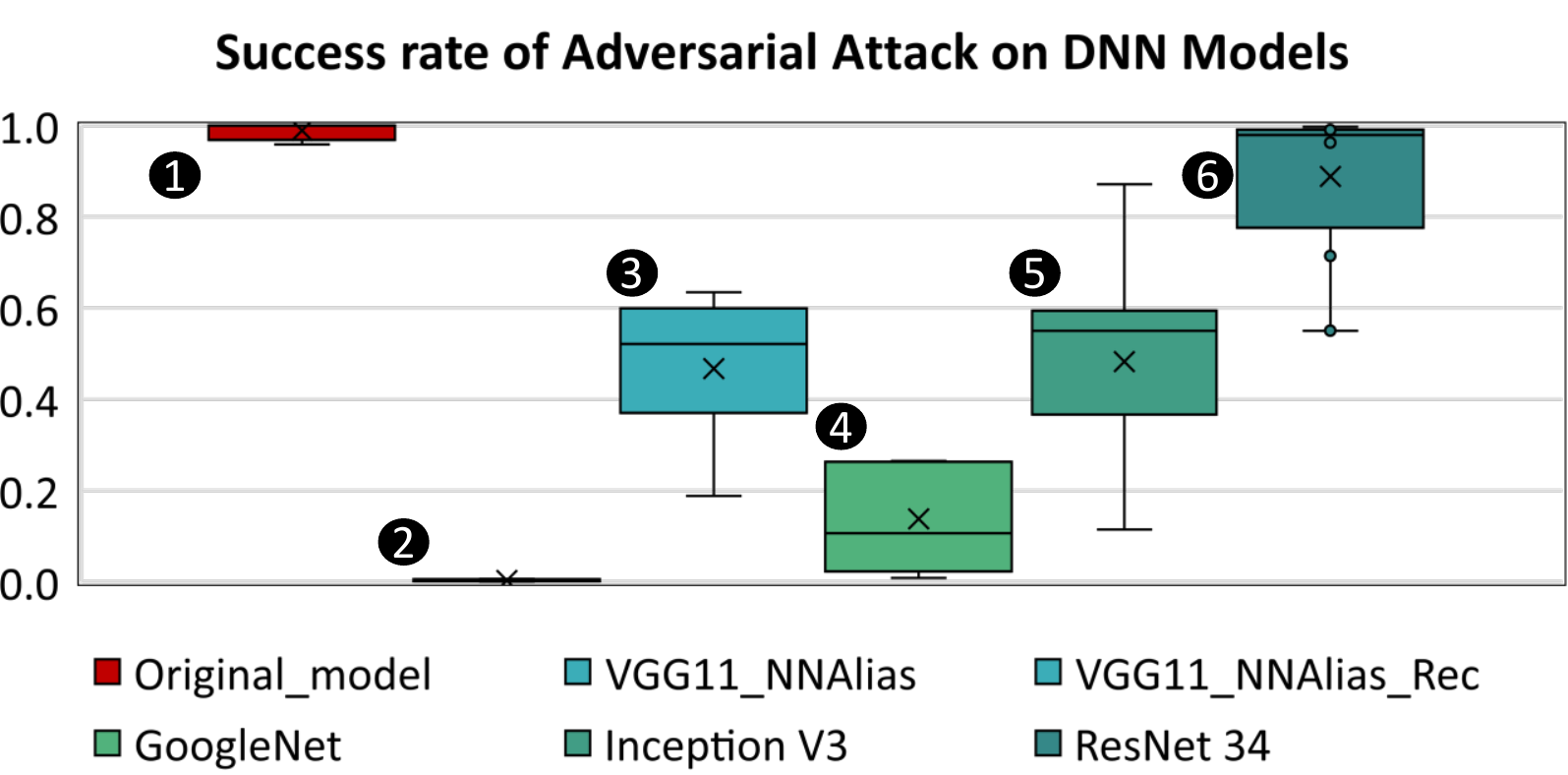}
 \vspace{-1.5em}
 \caption{Comparison of the average success rate of adversarial attack on VGG-11 model across various substitute models. Results show that only obfuscation was effective in mitigating the attack.}
 \vspace{-1em}
 \label{fig:Adv_result}
\end{figure}

 \textbf{1. Original Model:} 
 To validate the adversarial attack implementation, we tested adversarial samples generated from the unprotected model extracted by SCAS. The results, represented by \Circled{\scriptsize\textbf{1}} in Fig.~\ref{fig:Adv_result}, show a success rate of $98\%$ because the unprotected model closely resembles the original model.

 \textbf{2. Obfuscated and Recovered Models: }
 Next, we generate the adversarial samples  using the DNN obfuscated by DNN-Alias and test the target model. The results in \Circled{\scriptsize\textbf{2}} in Fig.~\ref{fig:Adv_result} show that the adversarial attack on DNN-Alias is unsuccessful (success rate $0.2\%$). Furthermore, we launch NeuroUnlock~\cite{NeuroUnlock} after the SCAS attack and we show (\Circled{\scriptsize\textbf{3}} in Fig.~\ref{fig:Adv_result}) that the success rate increases to $51\%$ on average. Although NeuroUnlock enhances the performance of the adversarial attack, the attack is still ineffective due to the errors in the de-obfuscation process.
 
 \textbf{3. Public DNN Families: } We report the success rate of the adversarial attack on target DNN when the adversarial samples are generated using standard DNN families. The success rate for GoogleNet \Circled{\scriptsize\textbf{4}} is $14\%$, Inception-V3 \Circled{\scriptsize\textbf{5}} is $48\%$, and ResNet-34 \Circled{\scriptsize\textbf{6}} is $88\%$. Since the attacker is unaware of the architecture of the target DNN to choose a similar DNN family, the results present that DNN-Alias successfully protects the DNN against SCAS-based adversarial attacks.

\vspace{-0.5em}
\subsection{Overhead Analysis}

The results in Fig. 17 show that while DNN-Alias on average, increases memory access time for both read ($13\%$) and write ($40\%$) operations, the computation latency decreases ($25\%$). Thus, memory access time presents the primary bottleneck for further increasing the obfuscation level. This observation opens up opportunities for future research on the optimization of obfuscation techniques through the use of efficient memory protocols.

\begin{figure}[!t]
 \centering
 \includegraphics[width=\linewidth]{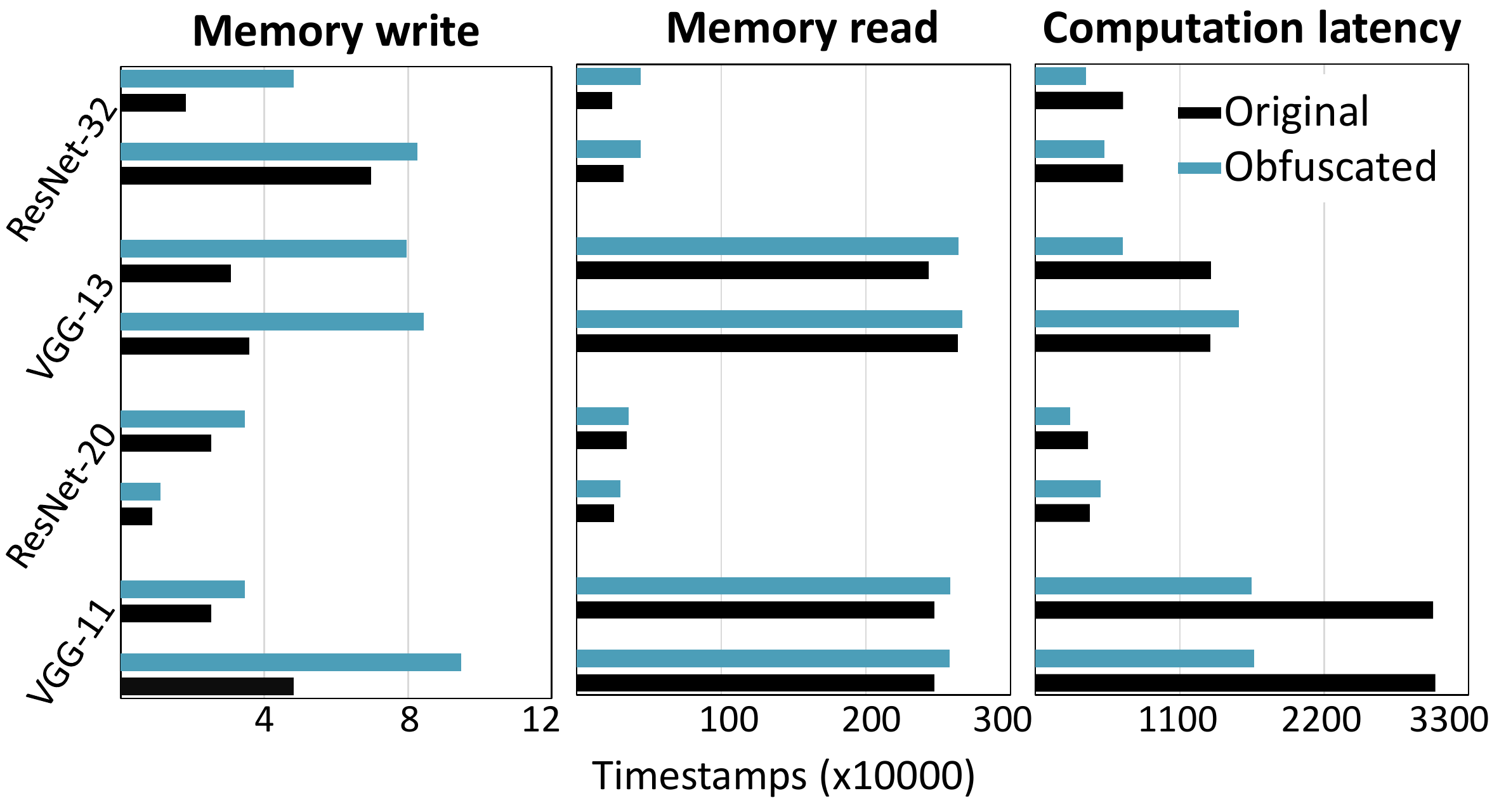}
\caption{Effect of DNN-Alias on memory access time and computation latency. The latency overhead of DNN-Alias is primarily attributed to increased memory accesses. Computational latency was reduced in most cases.}
 \vspace{-1em}
\label{fig:Profiled}
\end{figure}

%% file: Text/07_Conclusion.tex
\section{Conclusion}
\label{Conclusion}

In this paper, we present a novel obfuscation method called DNN-Alias to protect deep neural networks (DNNs) against side-channel attacks. Our proposed method forces all the layers in a DNN to have similar execution traces, making it difficult for attackers to differentiate between the layers and extract the architecture. DNN-Alias employs a genetic algorithm to find the best combination of layer obfuscation operations to maximize the security level while maintaining a user-specified latency overhead budget.

The effectiveness of DNN-Alias is demonstrated through experiments on various randomly generated and publicly available DNNs. We show that DNN-Alias can successfully prevent state-of-the-art side-channel architecture stealing attacks and adversarial attacks while preserving the original functionality of the DNNs. Our results highlight the potential of DNN-Alias as a generic and hardware-independent defense mechanism for DNNs against side-channel attacks.
 \vspace{-0.3em}